\documentclass{aa}

\usepackage{graphicx}
\usepackage{xcolor}
\usepackage{subcaption}
\usepackage{multirow}
\usepackage{placeins}
\usepackage{comment}

\usepackage[colorlinks,linkcolor=blue,citecolor=blue,urlcolor=blue]{hyperref}
\usepackage{txfonts}
\usepackage{booktabs}
\usepackage{orcidlink}

\newcommand*{\rom}[1]{\expandafter\@slowromancap\romannumeral #1@}

\begin{document}

   \title{Something green beneath the surface: The dynamical nature of fossil structures in IllustrisTNG-100}
   
   \titlerunning{The dynamical nature of fossil systems}

    \author{Mary Verdugo-Santos\corrauth{mary.verdugo@userena.cl}\orcidlink{0009-0000-2389-5837}\
          \inst{1}\and\
          Facundo A. G\'omez\orcidlink{0000-0003-4232-8584}\
          \inst{1}\and
          Diego Pallero\orcidlink{0000-0003-2150-1130}\
          \inst{2,5}\and\
          Franklin Ald\'as\orcidlink{0000-0002-9021-5114}\
          \inst{1,4,6}\and\
          Cristian A. Vega-Mart\'inez\orcidlink{0000-0002-4998-7606}\
          \inst{3}\
          }

    \authorrunning{Verdugo-Santos, M et al.} 

   \institute{Departamento de Astronom\'ia, Universidad de La Serena,
              Av. Raul Bitr\'an 1305, La Serena, Chile.
         \and
             Departamento de F\'isica, Universidad T\'ecnico Federico Santa Mar\'ia,
             Av. Espa\~na 1680, Valpara\'iso, Chile.
         \and
             Facultad de Ingenier\'ia y Arquitectura,
             Universidad Central de Chile,
             Av. Francisco de Aguirre 0405, La Serena, Chile.
         \and
             INAF -- Osservatorio Astronomico di Trieste,
             Via Tiepolo 11, 34143 Trieste, Italy.
         \and
             Núcleo Milenio de Galaxias (MINGAL), Chile.
         \and
             IFPU, Institute for Fundamental Physics of the Universe, via Beirut 2, 34151 Trieste, Italy.}

\abstract
{Fossil structures (FSs) have traditionally been considered dynamically relaxed end products of group evolution, characterized by a large magnitude gap ($\Delta m_{1,2} \geq 2$). However, recent observations and simulations suggested that this picture is incomplete, motivating a reassessment of their nature using cosmological simulations.}
{We investigated the dynamical relaxation state of FSs, the differences in their galaxy populations relative to non-fossil systems (non-FSs), and the assembly history of the two populations. We focused on the system dynamics and the evolution of galaxies within them.}
{Using IllustrisTNG-100, a state-of-the-art cosmological simulation, we selected 182 structures ($M_{200} > 10^{13}\,M_{\odot}$) at $z = 0$, classified them as FSs or non-FSs based on $\Delta m_{1,2}$ in the $r$ band. We tracked the $\Delta m_{1,2}$ evolution over 9\,Gyr and analyzed (1) the emergence of $\Delta m_{1,2}$ and its relation to the last significant accretion event onto the system, (2) the fraction of quenched galaxies (specific star formation rate $< 10^{-11}$\,yr$^{-1}$), (3) the distribution of galaxies in color--stellar mass space, and (4) the centroid shift in the gas--Brightest Structure Galaxy (BSG) as a dynamical proxy.}
{The magnitude gap in FSs is primarily driven by the absence of massive recent accretion: FSs exhibit significantly lower BSG-to-satellite stellar mass ratios ($\mu^{\rm{FS}}_{\star}$=0.17 vs. $\mu^{\rm{NFS}}_{\star}$=0.39) for the most massive satellite accreted within the last 6\,Gyr. FSs also host a more prominent red sequence and marginally higher quenched fractions than non-FSs. Furthermore, non-FSs show more prominent green-valley populations, which are associated with the secondary massive galaxy behind the small-magnitude gap. When we compared the gas--BSG centroid offsets, we found that both populations occupy an intermediate dynamical state (D$_{\rm BSG-CM} \approx$ 0.14--0.16 $R$/$R_{200}$) that is broadly consistent with unperturbed systems, but neither population has reached a relaxed state.}
{Our findings indicate that while the magnitude gap effectively identifies systems that have ceased major mergers in the last 3-6 Gyr, it is a poor proxy for their current global dynamical state, as reflected in the weak anticorrelation between $\Delta m_{1,2}$ and $D_{\rm BSG-CM}$. The FS and non-FS populations exhibit intermediate gas-BSG offsets ($D_{BSG-CM} \approx 0.15 R/R_{200}$) and fail to reach relaxation. This decoupling suggests that the magnitude gap traces the assembly history of massive components and not the overall stability of the intracluster medium.}


\maketitle
\nolinenumbers

\section{Introduction}
\label{sec:intro}

Galaxy associations span a wide mass spectrum that ranges from small groups with only a few members ($\approx 4$--$6$) to massive clusters hosting more than a thousand galaxies. 
These systems are natural outcomes of the hierarchical paradigm of structure formation \citep{WhiteandRees} and are observed across a wide range of redshifts. This allows us to study their evolution and the mechanisms by which galaxies transform their stellar populations, either through internal processes or via interactions with their environment \citep{Boselli2006}.
In particular, mergers within groups can lead to the formation of so-called fossil groups, which were first introduced by \citet{Ponman94} and are typically defined by a large magnitude gap ($\Delta m_{1,2} > 2$) between the brightest galaxy and its nearest companion within $0.5R_{200}$ \citep{Jones2003}.

The central galaxies in these systems are extremely luminous and resemble brightest cluster galaxies (BCGs), although they lack the dense environments typical of rich clusters. 
While originally Fossil groups are interpreted as end products of group evolution, recent studies extended the definition to more massive systems, including fossil clusters \citep{review}, highlighting the need to reassess their formation pathways.
The evolutionary paradigm of fossil structures (FSs, including groups and clusters) has undergone substantial revision over the past decade.
Recent theoretical and observational studies have proposed that FSs are transient phases susceptible to disruption by subsequent mergers or infall events \citep{Vonbendabeckmann2008}. 
In this context, studies such as \citet{Mendes2007} have proposed a dry merger scenario within compact groups, in which L*\footnote{Galaxies whose luminosity corresponds to the characteristic break in the Schechter luminosity function.} galaxies merge to form overluminous brightest group galaxies (BGGs), eventually leading to FSs. 
However, results from the Fossil Group Origins (FOGO) project \citep{fogo1} and simulations such as Illustris \citep[][hereafter K17]{Kundert2017} suggested a more complex picture. 
Not all FSs originate from compact groups, and systems with masses greater than \(10^{13}M_\odot\) often show substructures indicative of recent activity. 
These results support the idea that some FSs follow distinct evolutionary pathways driven by galaxy-galaxy mergers \citep{Grutzbauch2009} at later times rather than by early, isolated formation, highlighting the complexity of fossil system evolution. While \citealt{Vonbendabeckmann2008} proposed that the fossil phase might be temporary, \citealt{Cui2011} found that some systems have remained fossilized for over 9 Gyr, possibly due to insufficient time for defossilization.

The quantification of magnitude gaps and the classification of systems into fossil and non-fossil structures is typically performed within their central regions ($<0.5R_{\rm vir}$). 
Using this definition, K17 analyzed the Illustris simulation \citep{Illustris} to investigate the mass assembly of the BGG and the evolution of the magnitude gap in FSs compared to a non-FS control sample. 
They found that BGGs in FSs assembled relatively late, $\sim$3 Gyr ago on average, and that their growth was strongly affected by recent accretion events. 
This result indicates that the magnitude gap does not necessarily trace early-forming systems, but instead reflects the recent accretion history. 
However, these conclusions were limited by the baryonic physics model in Illustris and by the restricted halo mass range considered by K17 ($M_{200} < 10^{13-13.5}M_\odot$). 
A reassessment with more advanced cosmological simulations is therefore required.

Our study uses the IllustrisTNG cosmological simulation (hereafter TNG), which represents a significant improvement in physical modeling and resolution. 
The new dual-mode AGN feedback model, in which a kinetic wind mode operates at low accretion rates \citep{Weinberger2017}, and the more realistic implementation of baryonic processes \citep{NelsonTNG2018} motivate an updated characterization of FSs in their dynamical and stellar population properties. The updated modeling of the color bimodality \citep{PillepichTNG2018} enables us to extend the analysis of K17 to a wider halo mass range ($M_{200} > 10^{13}M_\odot$), thereby selecting FSs in group- and cluster-scale environments. We also explore the differences between fossil and non-fossil systems based on different radius selections ($r < 0.5R_{200}$ and $r < 1R_{200}$), and we examine the emergence and persistence of the magnitude gap ($\Delta m_{1,2}$) and its relation to the relaxation state. In addition, we quantify the timing of the last significant accretion event of each system, which allows us to compare the $\Delta m_{1,2}$ dynamical state and $\Delta m_{1,2}$accretion history relations.

This paper is organized as follows. In Sect. \ref{sec:methods} we describe our sample selection criteria ($M_\star > 10^{9.5}\,\mathrm{M_\odot}$ and $M_\mathrm{halo} > 10^{13}\,\mathrm{M_\odot}$), the definitions applied at two spatial scales ($0.5R_{200}$ and $1R_{200}$), the computation of gas--galaxy shifts, and the tracking of the magnitude gap $\Delta m_{1,2}$ across cosmic time. In Sect. \ref{sec:results} we present our main findings: (1) the magnitude gap in FSs emerged relatively recently, with FS and non-FS populations diverging at different epochs depending on halo mass, indicating that the gap traces recent accretion history rather than absolute age; (2) FSs experienced significantly fewer massive satellite accretion events in the recent past; (3) FSs host a more prominent red sequence and marginally higher quenched fractions, while non-FSs exhibit an overdensity of massive galaxies in the green valley, associated with the secondary galaxy sustaining the small magnitude gap; and (4) FSs and non-FSs both occupy an intermediate dynamical state. In Sect. \ref{sec:discussion} we discuss these results in the context of the overabundance of FSs in TNG-100, the implications for the quenching efficiency, and the difference between assembly history and dynamical relaxation. Finally, we summarize our conclusions in Sect. \ref{sec:conclusions}.

\section{Methods}
\label{sec:methods}
\subsection{The IllustrisTNG-100 simulation}

The IllustrisTNG project \citep[][]{SpringelTNG2018, PillepichTNG2018, NelsonTNG2018, NaimanTNG2018, MarinacciTNG2018} is a suite of large-volume cosmological magnetohydrodynamical simulations run with the moving-mesh code \textsc{AREPO} \citep{Springel2010}, adopting a flat $\Lambda$CDM cosmology with parameters from the Planck 2015 results \citep{planck2016}: $\Omega_m = 0.3089$, $\Omega_b = 0.0486$, $\Omega_\Lambda = 0.6911$, $\sigma_8 = 0.8159$, $n_s = 0.9667$, and $h = 0.6774$. The project includes three flagship runs of increasing volume and decreasing resolution: TNG-50, TNG-100, and TNG-300, with box sizes of $\sim 50$, $100$, and $300\,\mathrm{Mpc}$ on a side, respectively. We used the TNG-100 run, which provides an optimal compromise between cosmological volume and mass resolution for studying group and cluster-scale halos.

As a successor to the original Illustris simulation \citep{Illustris, Illustris1, Illustris2, Illustris3}, TNG-100 updates the galaxy formation model by incorporating a dual AGN feedback mechanism: thermal feedback for high-accretion black holes and kinetic feedback for low-accretion ones. 
This directly improves the efficiency of quenching in massive galaxies, and together with the introduction of magnetohydrodynamics, it yields more realistic star formation rates, morphologies, and gas content compared to its predecessor. 
These improvements provide a more accurate modeling of the ICM properties within groups and clusters, which is directly relevant for characterizing the galaxy populations we studied \citep{PillepichTNG2018, NelsonTNG2018}.

The IllustrisTNG model reproduces a wide range of observed galaxy properties, including the stellar mass function, the stellar-to-halo mass relation, and the cosmic star formation rate density at $z \leq 10$. 
Group catalogs are generated using the FoF \citep{Davis1985} and \textsc{Subfind} \citep{Springel2001} algorithms from $z=127$ to $z=0$. The mass resolution of TNG-100 ($m_{\rm b} = 1.4 \times 10^6\,M_{\odot}$, $m_{\rm DM} = 7.5 \times 10^6\,M_{\odot}$) is sufficient to resolve individual galaxies within groups and clusters, enabling reliable measurements of stellar masses, star formation rates, and photometric properties, as well as tracing the evolutionary pathways of FS progenitors over the past $\sim$9\,Gyr.

\begin{figure}
    \centering
    \includegraphics[width=1\linewidth]{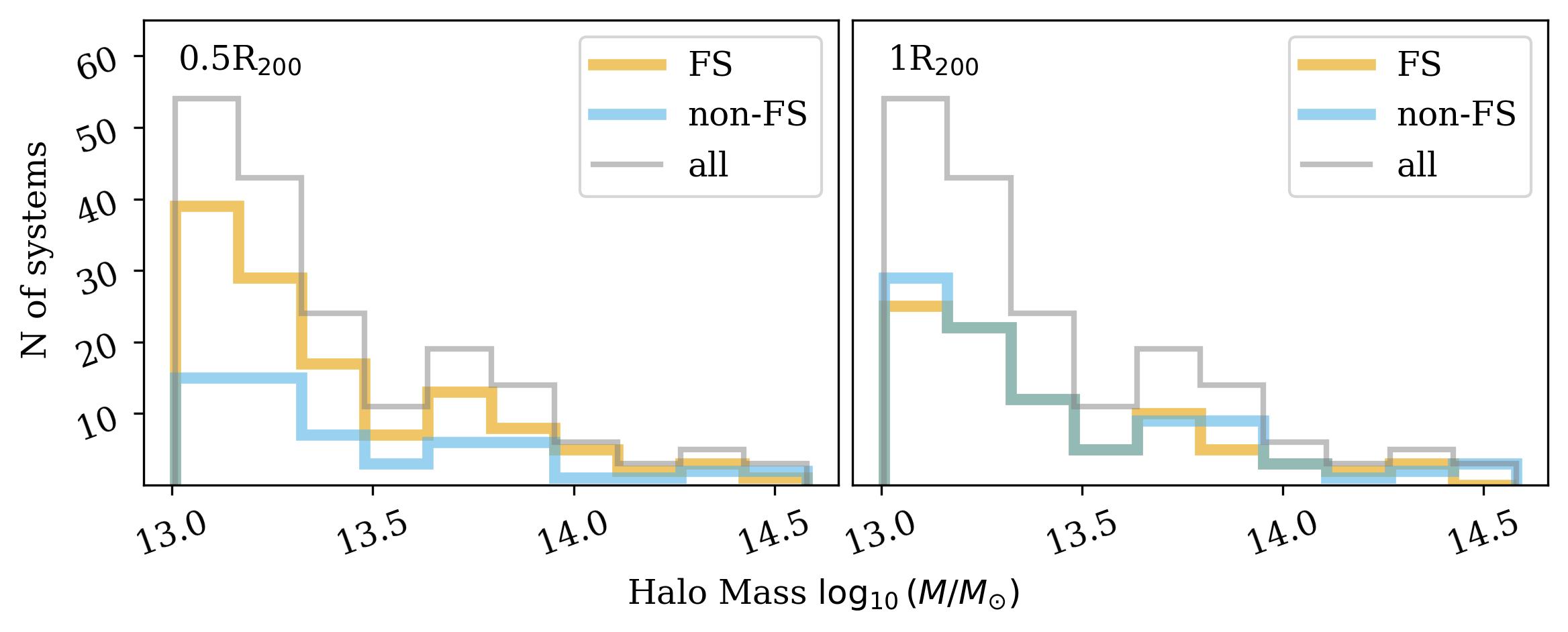}
    \caption{Halo mass distribution of our TNG-100 sample at $z=0$, classified by the fossil state at $0.5\,R_{200}$ (left panel) and $1\,R_{200}$ (right panel). The orange and blue histograms represent the FS and non-FS populations, respectively. The two populations span a mass range of $\log_{10}(M_{200}/M_{\odot}) \geq 13$.}
    \label{fig:halo_masses}
\end{figure}

\subsection{Halo mass and stellar mass selection criteria}

\label{sec:Halo_selec}

Groups and clusters of galaxies were selected from the publicly available catalogs of the IllustrisTNG-100 simulation. 
We considered as galaxies all subhalos with stellar masses $M_{\star} > 10^{9.5} M_{\odot}$, which corresponds to the resolution limit at which TNG-100 reliably resolves global galaxy properties \citep{PillepichTNG2018, NelsonTNG2018}.
This threshold ensured that each galaxy was represented by 2000 stellar particles at least, providing a sufficient numerical resolution for measurements of stellar masses, star formation rates, and photometric properties.
Below this limit, numerical artifacts and stochastic effects significantly affect the measured galaxy properties.
Following the mass criterion established by \citet{Jones2003}, we selected structures with halo masses $M_{\mathrm{halo}} > 10^{13} M_{\odot}$, according to the friends-of-friends catalog. 
No upper mass limit was imposed, which allowed us to extend our sample from galaxy groups into the cluster mass regime.

To investigate potential mass-dependent trends in the selected structures properties, we further subdivided our sample into two mass bins: $10^{13} \leq M_{\mathrm{halo}}/M_{\odot} < 10^{13.5}$ and $M_{\mathrm{halo}}/M_{\odot} \geq 10^{13.5}$. 
This division allowed us to examine whether structural differences become more pronounced with increasing halo mass and to identify any systematic variations in the fossil characteristics at different mass scales. 
The mass distribution of the $z=0$ selected structures is shown in Figure \ref{fig:halo_masses}.

In order to increase the number of structures, following \citet[][hereafter A25]{Aldas2024} we further expended our sample by selecting structures at different redshifts, $z$. We assumed that the dynamical state of the structures changes over time. In particular, we selected structures a $z=0$, 0.2 and 0.4.

\begin{figure*}
    \centering
    \includegraphics[width=1\linewidth]{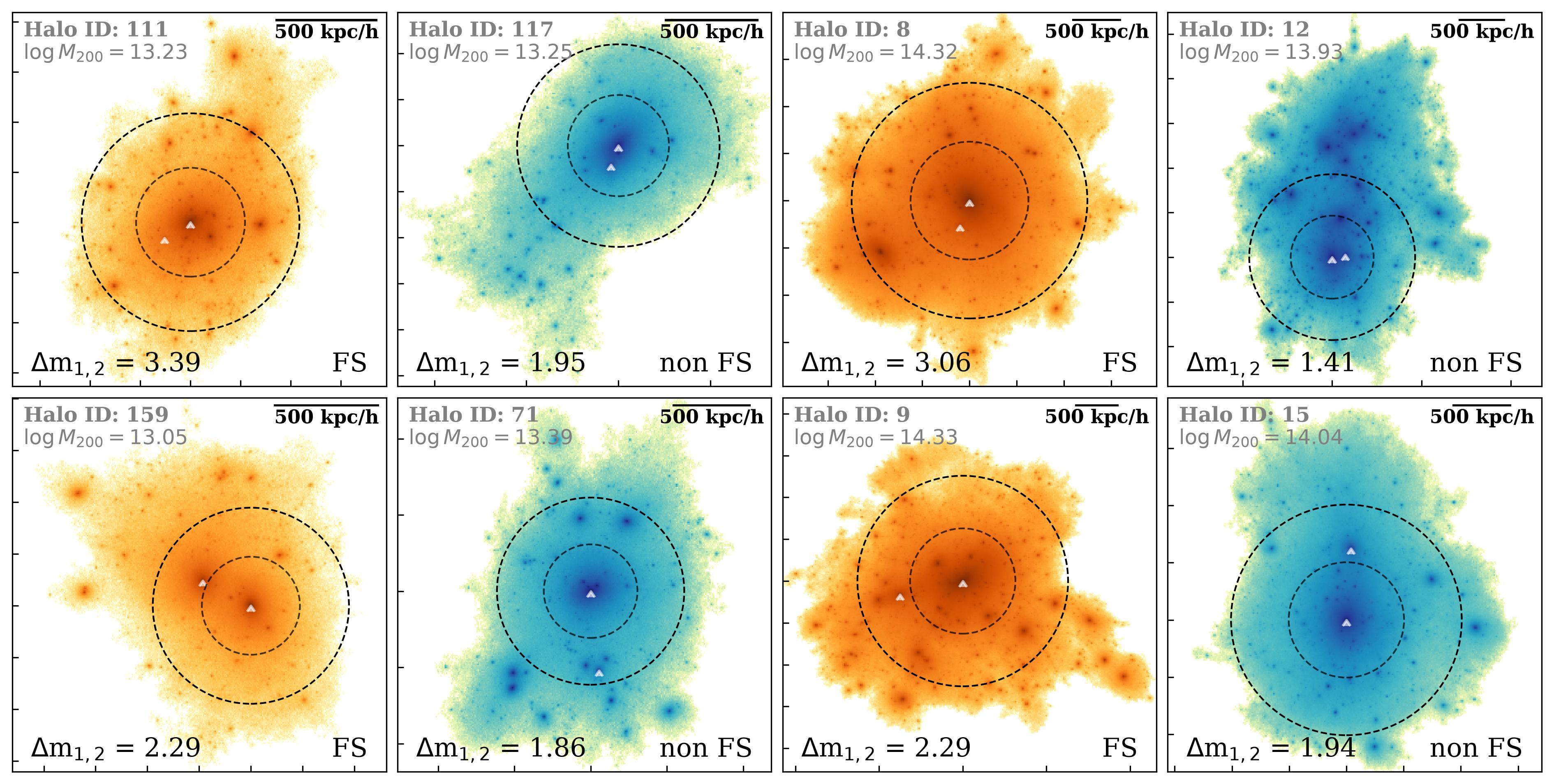}
    \caption{Stellar particle distributions of representative FS (orange, first and third columns) and non-FS (blue, second and fourth columns) systems from TNG-100 at $z=0$, shown in two halo mass ranges. Left two columns: Structures with $10^{13} \leq M_{200}/M_\odot < 10^{13.5}$. Right two columns: Structures with $M_{200} \geq 10^{13.5}\,M_\odot$. The top and bottom rows show selected halos based on the $0.5\,R_{200}$ and $R_{200}$ criterion, respectively. The dashed black circles indicate $0.5\,R_{200}$ and $R_{200}$. The white triangles mark the two brightest galaxies within the aperture adopted in each row whose magnitude difference defines the fossil classification criterion ($\Delta m_{1,2} \geq 2$).}
    \label{fig:examples}
\end{figure*}

\begin{table}

\caption{Number of systems per sample at different redshifts and apertures.}
\label{tab:sample}
\centering
\begin{tabular}{l rr @{\hspace{1em}} rr}
\hline\hline
\noalign{\smallskip}
& \multicolumn{2}{c}{$z = 0$}
& \multicolumn{2}{c}{$0 < z \leq 0.4$} \\
\cmidrule(lr){2-3}\cmidrule(lr){4-5}
& \multicolumn{4}{c}{$\log(M_{200}/M_{\odot})$} \\
\cmidrule(lr){2-5}
Sample
  & $[13,\,13.5)$
  & ${\geq}\,13.5$
  & $[13,\,13.5)$
  & ${\geq}\,13.5$ \\
\noalign{\smallskip}
\hline
\hline
\noalign{\smallskip}
\multicolumn{5}{l}{$0.5\,R_{200}$} \\
\noalign{\smallskip}
FS  &  85 & 39 & 160 & 73 \\
NFS &  37 & 21 &  112 & 33 \\
\noalign{\smallskip}
\noalign{\smallskip}
\hline
\noalign{\smallskip}
\multicolumn{5}{l}{$1\,R_{200}$} \\
\noalign{\smallskip}
FS  &  59 & 28 & 96 & 42 \\
NFS &  63 & 32 & 154 & 64 \\
\noalign{\smallskip}
\hline
\end{tabular}
\tablefoot{Within the FS sample, systems at $z=0$ that lack a sufficiently bright neighboring galaxy for computing the magnitude gap are assigned an effectively infinite gap; they are classified as isolated fossils and included in the FS totals.}
\label{tabla}

\end{table}

\subsection{Fossil classification via r-band magnitude gap}
\label{sec:fsc}

We classified galaxy groups and clusters as fossil and non-fossil structures based on the quantity known as the magnitude gap, $\Delta m_{1,2}$. 
It is calculated as the difference in r-band magnitude between the brightest galaxy (the brightest structure galaxy, BCG/BGG, hereafter BSG) and the second brightest galaxy (the brightest satellite) within a specified projected radius: $\Delta m_{1,2} = m_2 - m_1$.
A large magnitude gap ($\Delta m_{1,2} \geq 2$ mag) indicates that the central galaxy is significantly more luminous than any other galaxy in the system, suggesting either preferential growth of the central galaxy through hierarchical mergers \citep{DeLucia&Blaizot2007, Dariush2010} or depletion of the intermediate-luminosity satellite population \citep{Raouf2014, FOGO52015}.
Beyond its role as a selection criterion, several studies have proposed the magnitude gap as a proxy for the dynamical state, with larger gaps potentially indicating more dynamically relaxed systems that have experienced sufficient time for the luminosity hierarchy to develop \citep{Dariush2007, Gozaliasl2014,FOGO11}. However, this interpretation remains debated, as other works suggested that large gaps can also arise in dynamically young systems through stochastic assembly histories (\citealt{Cui2011}, K17).

Following the canonical definition by \citet{Jones2003, Sun2004, FOGO22012}, we classified structures at $z=0$ with $\Delta m_{1,2} \geq 2$ mag as FSs, while those with $\Delta m_{1,2} < 2$ mag constitute our non-fossil structure (non-FS) control sample.
This classification was performed independently at two radial scales: within $0.5R_{200}$ (the traditional definition) and within $1R_{200}$. Figure \ref{fig:examples} shows examples of selected FSs and non-FSs at z=0. 

After the Fossil and non-Fossil structures at $z=0$ were identified, we tracked their evolution across time using the merger trees provided by TNG-100 (\texttt{LHaloTree}, \citealt{Springel2005}). Within each FoF halo, gravitationally bound substructures were identified using the algorithm \textsc{Subfind} \citep{Springel2001}. The central galaxy of each group corresponds to the most gravitationally bound subhalo within the FoF halo (\cite{Springel2001}).
Since the brightness of satellite galaxies evolves with time, the magnitude gap is not a fixed quantity. Structures can transition between fossil and non-fossil states as their satellite populations change. To capture this evolution, we systematically recomputed $\Delta m_{1,2}$ at each simulation snapshot, classifying each structure's fossil status iteratively from $z = 0$ back to $z \sim 1$.

Table \ref{tabla} shows the results of our classification procedure. In the group  mass range ($10^{13}$ -- $10^{13.5}$ M$_{\odot}$), 62.2\% and 41.7\% of the selected structures are fossil when they were selected  based on the 0.5$R_{200}$ and $R_{200}$ criteria. Although less prominent, as in K17, we found an overabundance of FSs with respect to observations \citep[e.g.][]{SDSS}. K17 reported a relative abundance of 80\% of FSs with respect to non-FSs. The  smaller fraction of FSs we found is  associated with the better modeling of central galaxies implemented in IllustrisTNG, which prevents their excessive mass growth. We further explore this in Section \ref{sec:analysis:magnitudegap}. Other aspects discussed in K17 that contribute to this overabundance, such as over-merging of satellite galaxies or the lack of X-ray luminosity criterion to select structures, will be analyzed in a follow-up study.

\subsection{Specific star formation rate and quenching criteria}

The star formation rate (SFR) of the galaxies in each structure was extracted from the publicly available catalog. This property was calculated by adding at $z=0$ the instantaneous SFR of the gas cell bound to each subhalo. To obtain the specific star formation rate (sSFR), we divided the SFR in the corresponding stellar mass of each galaxy.

To compare the evolutionary state of galaxies within the groups and clusters, we used the concept of a quenched galaxy. We followed the definition by \citealt{Wetzel2012} of a quenched galaxy at $z=0$ as a galaxy with an sSFR below 10$^{-11}$ yr$^{-1}$. 
We then classified galaxies within the structures as either quenched or non-quenched, and we then computed the fraction of quenched galaxies:
This same threshold was adopted in previous studies characterizing the star-forming and quenched populations of the IllustrisTNG simulations \citep{Donnari2019}, which ensures consistency with previous works,

\begin{equation}
    \hspace{80pt} f_{\rm Q} = \frac{N_\textnormal{quenched galaxies}}{N_\textnormal{total galaxies}}    .
\end{equation}

This method allowed us to estimate the percentage of quenched galaxies within the selected FSs and non-FSs, and it thus provided insight into their different evolutionary paths and the effect produced in the galaxy population.

\subsection{Classification of the dynamical state}
\label{sec:methods:dynamical}

To assess the dynamical state of groups and clusters in our sample, we employed a morphological indicator based on the distance between the position of the BSG and the gas center-of-mass (CoM) of the entire structure, $D_{\rm BSG-CM}$.
Following \citet{DeLuca2021}, we used the BSG position as a tracer of the underlying matter distribution because it is expected to coincide with the density peak in relaxed structures. On the other hand, we used the position of the CoM of the gas as a proxy for the Sunyaev-Zeldovich (SZ) effect centroid (A25).
The physical basis of this indicator lies in the different response of the collisionless and collisional components to a merger. The BSG is a collisionless tracer, and it lies at the bottom of the potential well, which is dominated by the dark matter distribution \citep{Vandenbosh2005}. The gas, by contrast, is collisional and experiences ram pressure and hydrodynamical drag during mergers, which decouples it from the collisionless components. The different responses separate the BSG and the gas centroid in interacting systems, whereas in relaxed structures, they both remain closely coincident near the potential minimum.

We calculated the gas CoM using all gas particles associated with the friends-of-friends (FoF) group of each halo. Because the FoF  selection is not restricted to a fixed aperture, gas particles beyond $1\,R_{200}$ can belong to a neighboring structure that interacts with the halo.

Following the classification scheme of A25, we considered a structure dynamically disturbed when the offset between the BSG and the center of mass of the gas satisfied $D_{\rm BSG-CM} > 0.4\,R_{200}$, where $R_{200}$ is the radius within which the mean density is 200 times the critical density of the Universe.
This threshold allowed us to distinguish between relaxed systems, where the BSG traces the gravitational potential minimum, and disturbed systems that have likely experienced recent mergers or accretion events. This procedure was applied uniformly to FSs and non-FSs at $z=0$, $z=0.2$, and $z=0.4$. By comparing FSs and non-FSs, we aimed to determine whether fossil structures exhibit statistically smaller shifts, which would be consistent with the scenario of a more relaxed intrastructure medium.

\section{Analysis and results}
\label{sec:results}

\begin{figure*}
    \centering
    \includegraphics[width=1\linewidth]{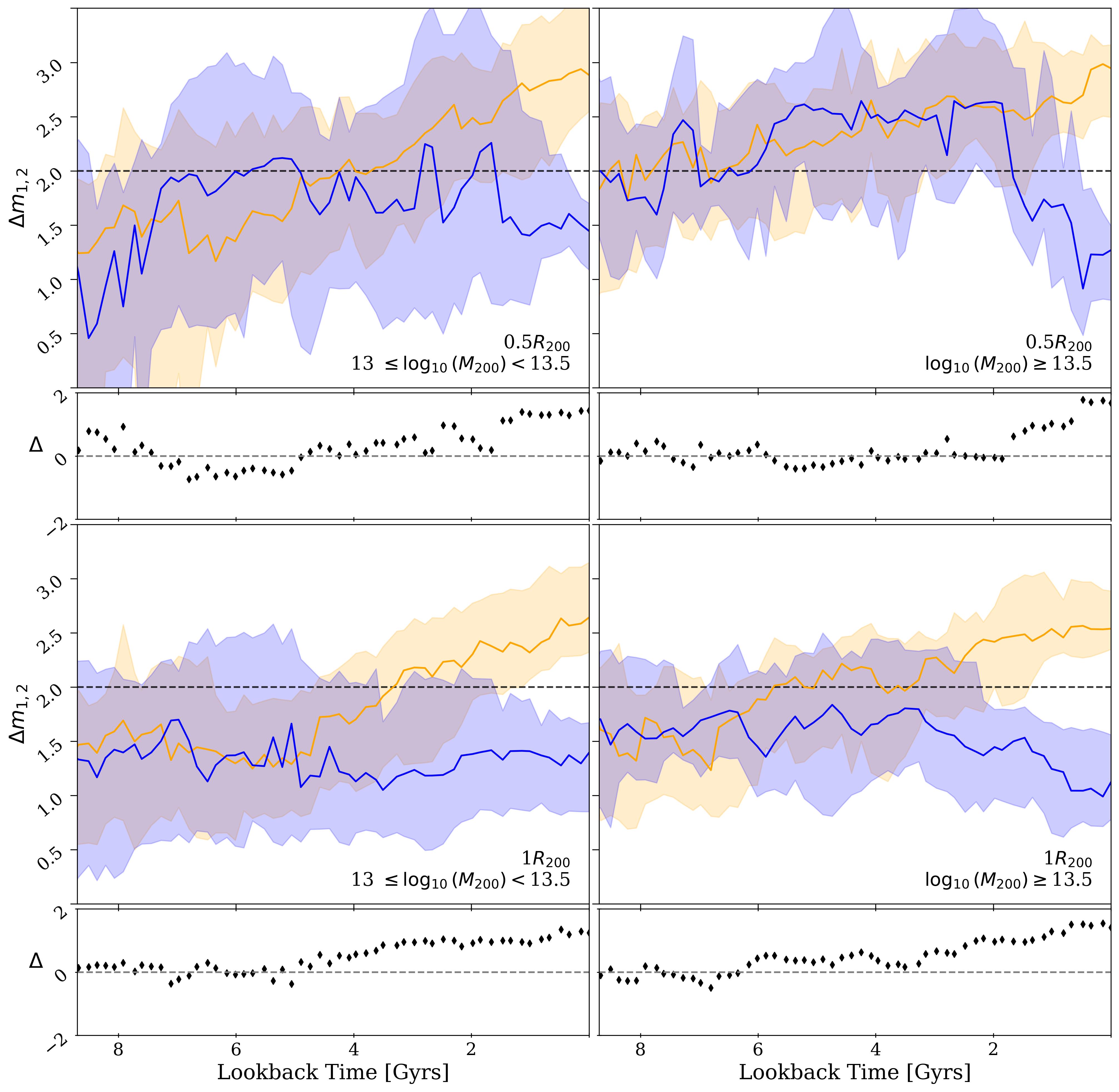}
    \caption{Time evolution of the median magnitude gap $\Delta m_{1,2}$ over a lookback time of $\sim 8.5$\,Gyr for systems classified at $0.5\,R_{200}$ and $1\,R_{200}$ (upper and lower main panels, respectively), each accompanied by a residual subpanel showing $\Delta$ (FS $-$ non-FS). The left and right columns correspond to halo mass bins of $13 \leq \log_{10}(M_{200}/M_{\odot}) < 13.5$ and $\log_{10}(M_{200}/M_{\odot}) \geq 13.5$, respectively. The orange and blue lines show the median $\Delta m_{1,2}$ for FS and non-FS populations, and the shaded regions indicate the interquartile range. The dashed horizontal line marks the fossil threshold $\Delta m_{1,2} = 2$. The sample sizes decrease with lookback time because the merger tree is incomplete at high redshift.}
    \label{fig:magnitudegap}
\end{figure*}

\subsection{Evolution of the magnitude gap: Emergence and persistence}
\label{sec:analysis:magnitudegap}

The evolution of the magnitude gap over cosmic time can allow us to trace when FSs and non-FSs, as populations, begin to diverge. Various authors have therefore proposed magnitude gaps as reliable indicators of the structure dynamical state \citep{Goldenmarx2025, Yang2025}. FSs are often considered dynamically relaxed since they exhibit large $\Delta m_{1,2}$. However, this metric can easily be affected by a recent accretion event that might not necessarily globally drive the system into an unrelaxed state. Furthermore, an unrelaxed structure that has recently undergone mass accretion can exhibit a large $\Delta m_{1,2}$ due to the relatively rapid tidal disruption of the most massive group member in the low-mass regime and preprocessing during group accretion in the high-mass regime \citep{Khalid2025}.

In Figure \ref{fig:magnitudegap} we show the time evolution of the median magnitude gap of the present-day FS and non-FS populations. This figure reveals distinct evolutionary patterns in different mass ranges. We first focus on the classification based on galaxy members located within $0.5R_{200}$. For structures with masses between $10^{13}$ and $10^{13.5}\,M_{\odot}$ (top left panel), the selected FSs began transitioning to fossil status approximately 4 Gyr ago, while non-FSs in the same mass range consistently maintained their non-fossil state on average.
The behavior of higher-mass structures ($>10^{13.5}\,M_{\odot}$, top right panel) is more complex. The non-FS population apparently experienced a fossil phase between 6 and 2 Gyr ago before transitioning to the non-fossil status within the last 2 Gyr.
In contrast, the FS population shows a consistent, mild rise toward the present day.

When structures are classified based on the satellite population located within R$_{200}$ (bottom panels), the evolutionary timescales differ slightly. 
For lower-mass structures (left bottom panel), the differences in the $\Delta m_{1,2}$ of present-day FSs and non-FSs begin to arise $\approx 5$ Gyr ago, and the FSs emerge within the last 3 Gyr.  
The corresponding non-FS population maintains its state throughout the entire 9 Gyr we analyzed. This is a consequence of the larger radial distance we considered to compute the magnitude gap. 
Higher-mass structures (right bottom panel) show significant divergence at z=0, with populations separating approximately 6 Gyr ago and FS emergence before 4 Gyr ago.
Critically, after the FS population crosses the threshold of $\Delta m_{1,2} \geq 2$ mag, it does not revert to the non-fossil state in either mass regime. 

The transition of non-FSs to FSs, as indicated by the magnitude gap, suggests that with respect to FSs, the non-FS population has experienced  significant accretion of more massive structures in the recent past. To quantify this, we defined the parameter  $\mu_{\star}$ as the stellar mass ratio of the most massive satellite galaxy to have crossed R$_{200}$ within the last 6 Gyr and the central galaxy of the host halo at the time when the satellites cross R$_{200}$. We adopted the stellar mass and not the total mass because the criterion for fossil system classification is based on the optical luminosity, which is directly tied to the stellar content of the galaxies.
The median $\mu_{\star}$ for the FS and non-FS samples is 0.17 and 0.39, respectively. The lower median $\mu_{\star}$ for the FS population shows that the interactions experienced by non-FSs within the last 6 Gyr have indeed been stronger and more significant than those experienced by the FS population. 

Using the Illustris simulation, K17 investigated the evolution of the magnitude gap of structures in the group mass range (M$_{200}$ = 10$^{13-13.5}$ M$_\odot$). Their results supported the idea of a fossil phase \cite{Vonbendabeckmann2008}. This means that for galaxies within $R_{200}$, FSs identified at z=0 were non-FSs at high redshift, while $z=0$ non-FSs were previously FSs. Moreover, the authors reported that all $z=0$ FSs were previously non-FSs, and vice versa. For the same mass range, our IllustrisTNG results agree on the present-day FS population. However, in contrast to K17, we found that the non-FS population has remained in this stage during the last 8 Gyr of evolution in general.
Interestingly, K17 reported that the difference between FS and non-FS $\Delta m_{1,2}$ starts to diverge strongly at $t_{\rm lb}$ $\approx 3$ Gyr, whereas we found this divergence to start slightly earlier, at $t_{\rm lb}$ {$\approx 5$ Gyr (bottom left panel of Fig.\ref{fig:magnitudegap}). For higher-mass structures ($M_{200} > 10^{13.5}$), we found a very similar behavior, but the divergence in the $\Delta m_{1,2}$ takes place at an even earlier time: $t_{\rm lb} \approx 5$ Gyr.

The discrepancy with K17 stems from fundamental differences in feedback mechanisms between the original Illustris-100 and TNG-100 simulations. The original Illustris simulation quenched star formation inefficiently in massive halos, resulting in a weak red sequence and an overabundance of massive, blue galaxies. Conversely, TNG-100 incorporates a kinetic wind-feedback mode in the low-accretion state. This mechanism effectively suppresses star formation in galaxies above a characteristic mass $M_* \sim 10^{10.5} M_{\odot}$. This quenching mechanism directly alters the spectral energy distribution and luminosity of massive galaxies, particularly in the r band ($M_r$), which traces older stellar populations  \cite{Weinberger2017, Pillepich2018model, NelsonTNG2018}. By preventing the runaway stellar mass growth seen in the original Illustris, TNG-100 yields Brightest Cluster Galaxies (BCGs) that are less luminous than their Illustris counterparts and consistent with the passive red population. Because the magnitude gap is defined by the luminosity contrast between the first- and second-ranked galaxies, the specific suppression of BSG $M_r$ luminosity in TNG, driven by low-state kinetic AGN feedback, primarily drives the divergent gap evolution found here. Our results, based on the improved physical model, suggest that the magnitude gap in FSs emerged $\sim$4–5 Gyr ago, when FSs and non-FSs diverged.

\subsection{Global galaxy properties in FSs and non-FSs}

Galaxy properties serve as fossil records of their host system's formation history. In this context, a fundamental question is whether the specific evolutionary pathway of FSs, characterized by early assembly and a lack of recent major mergers, leaves a distinct imprint on their satellite populations compared to non-fossil systems. If the extended isolation of FSs plays a dominant role, we expect systematic differences in properties such as enhanced quenched fractions or distinct color distributions. Alternatively, if galaxy evolution is primarily governed by environmental processes common to all massive halos (e.g., ram-pressure stripping and tidal interactions), FSs and non-FSs are both expected to exhibit comparable galaxy populations.

To test these scenarios, we examined the present-day color-magnitude distributions, specific star formation rates (sSFR), and fraction of quenched galaxies in both samples. We investigated whether galaxies in FSs and non-FSs follow the established correlations between sSFR and color \citep{Baldry2004} and if they maintain the characteristic galaxy color bimodality \citep{Kauffmann2003}. Additionally, we assessed whether the enhanced quenched population expected in FSs due to their potentially earlier formation is observationally significant compared to non-fossil systems.

\subsubsection{Quenched galaxy fractions in FSs and non-FSs}
\label{sec:fraction_of_quenched_galaxies}
As discussed in Sec. \ref{sec:methods}, we considered a galaxy as quenched if its sSFR $< 10^{-11}$. In Figure \ref{fig:QF_bins} we show the cumulative distribution function of galaxies found within different structures (gray histograms). Top and bottom panels show results obtained for galaxies selected within $0.5R_{200}$ and $R_{200}$, respectively. First, low-mass structures (left panels) show a great diversity in the fraction of quenched galaxies, $f_{\rm Q}$. We find structures that are dominated by star-forming galaxies ($f_{\rm Q} \approx 20\%$) and structures with a fully quenched galaxy population. The median is $f_{\rm Q} = 55\% $.  Conversely, higher-mass structures (right panels) are typically characterized by higher values of $f_{\rm Q}$. The median is $f_{\rm Q} = 77\% $. 

The top panels show the results obtained when the samples are divided into FSs (orange) and non-FSs (light blue) according to the $0.5R_{200}$ criterion (see Sec. \ref{sec:fsc}). For the low-mass subsample, non-FSs show a slightly higher fraction of structures with low $f_{\rm Q}$ than their FS counterparts. The cumulative distribution function (CDF) of FSs rises more sharply at intermediate values. This is better highlighted by the difference between the two distributions in the panels in the second row, where the maximum difference is $\Delta \approx  20\%$. A Kolmogorov-Smirnov (KS) test, performed to explore whether we can rule out that both functions are drawn from the same parent distribution, showed that this hypothesis cannot be rejected, however.
When analyzing the higher-mass sample (top right panel), we found a more pronounced discrepancy. FSs tend to exhibit a larger fraction of quenched galaxies than their non-fossil counterparts, with a maximum difference of $\Delta \approx  30\%$. Nonetheless, because these high-mass bins only contain a few structures, the difference between the two CDFs is not sufficient to formally rule out a common parent distribution at the 95\% confidence level (see table \ref{tab:ks_fq}). 
The systematic nature of this offset in most of the $f_{\rm Q}$ range might suggests an underlying physical trend that might be linked to the advanced evolutionary state of fossil halos. 

In the bottom panels of Figure \ref{fig:QF_bins}, we repeat the analysis, but for galaxies within a larger radius, that is, $1R_{200}$. We recall that the classification of a system as FS or non-FS depends on the radius. A system labeled as fossil within $0.5R_{200}$ can be reclassified as non-fossil when a bright secondary is located between $0.5$ and $1R_{200}$.  Interestingly, when a larger volume is probed, the differences between FSs and non-FSs become less pronounced in both mass bins. For the low-mass sample (bottom left panel), the maximum difference between the CDFs decreases to $\Delta \approx 16\%$, with a KS test p-value of $0.65$. A similar trend is observed in the high-mass subsample (bottom right panel), where the previously noted discrepancy ($\Delta \approx 30\%$ at $0.5R_{200}$) drops to $\Delta \approx 16\%$, and the mean quenching fractions become nearly identical ($\langle f_{\rm Q} \rangle \approx 73\%$ for both groups). The fact that the differences are more prominent when structures are classified based on the galaxy populations located in the inner regions suggests that aside from relatively massive recent accretion events (which would break the fossil condition), the environments of FSs and non-FSs are equally efficient in shaping the global properties of these structures. The higher number of galaxies selected at larger radii appears to dilute the signal, as the inclusion of a large population of already-processed galaxies, leading to nearly identical global quenching fractions.

\begin{figure}
    \centering
    \includegraphics[width=1\linewidth]{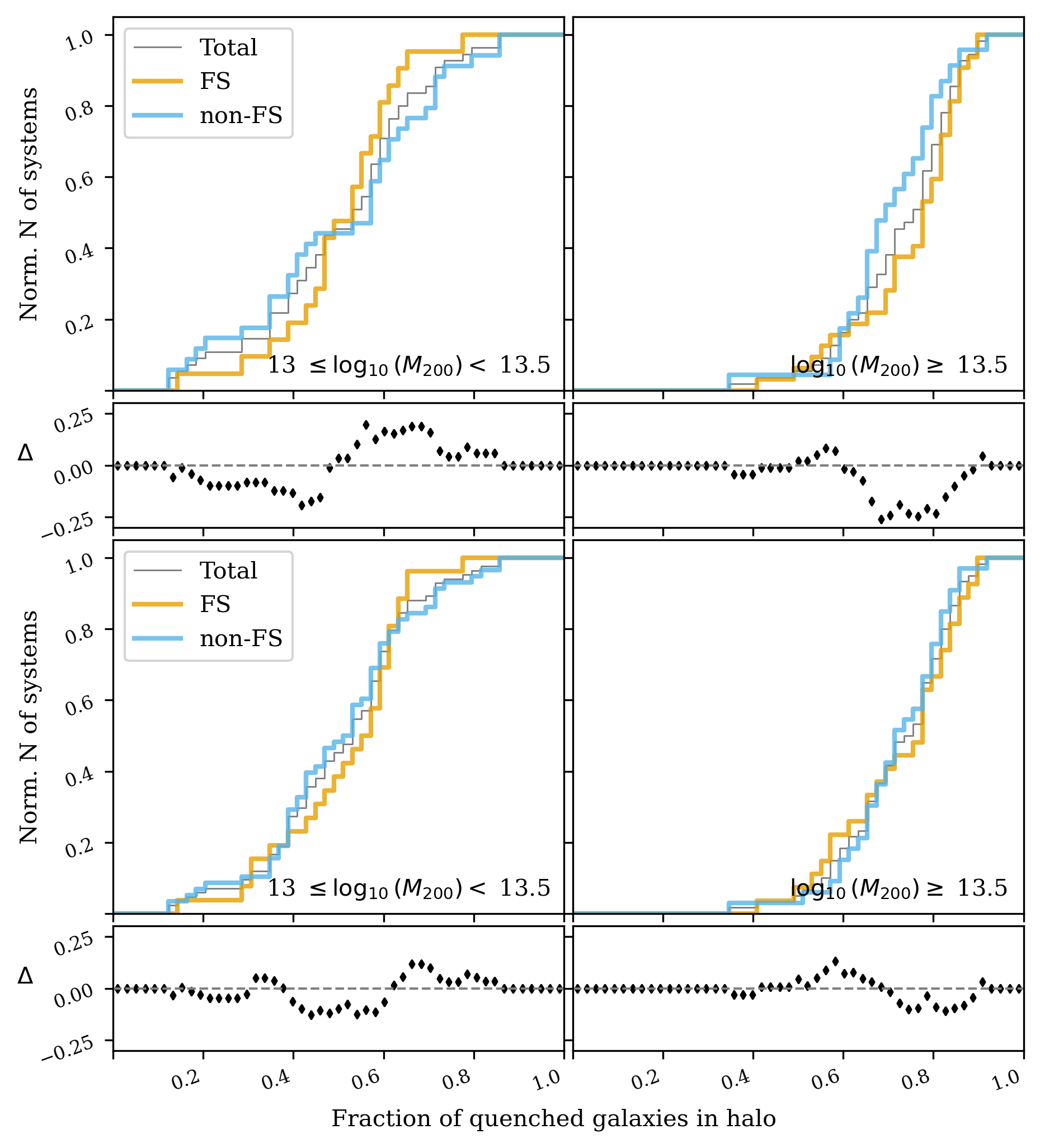}
    \caption{Cumulative distribution functions  of the quenched galaxy fraction ($f_{\rm Q}$, defined as sSFR $< 10^{-11}$\,yr$^{-1}$) for FS (orange), non-FS (blue), and total (gray) populations. The systems are classified at $0.5\,R_{200}$ and $1\,R_{200}$ (upper and lower main panels, respectively), each accompanied by a residual subpanel showing $\Delta$ (FS $-$ non-FS). The left and right columns correspond to halo mass bins $13 \leq \log_{10}(M_{200}/M_{\odot}) < 13.5$ and $\log_{10}(M_{200}/M_{\odot}) \geq 13.5$, respectively.}
    \label{fig:QF_bins}
\end{figure}

\begin{table}
\caption{Two-sample Kolmogorov--Smirnov tests comparing the $f_{\rm Q}$
distributions of FS and non-FS systems.}
\label{tab:ks_fq}
\centering
\setlength{\tabcolsep}{8pt}
\renewcommand{\arraystretch}{1.15}
\begin{tabular}{l c c}
\toprule\toprule
$\log_{10}(M_{200}/M_{\odot})$ & $D_{\rm KS}$ & $p$-value \\
\midrule
\addlinespace[1pt]
\multicolumn{3}{c}{Classification within $0.5\,R_{200}$} \\
\addlinespace[2pt]
$[13.0,\,13.5)$ & 0.214 & 0.51 \\
$\geq 13.5$     & 0.303 & 0.13 \\
\addlinespace[3pt]
\midrule
\addlinespace[1pt]
\multicolumn{3}{c}{Classification within $1\,R_{200}$} \\
\addlinespace[2pt]
$[13.0,\,13.5)$ & 0.163 & 0.65 \\
$\geq 13.5$     & 0.162 & 0.77 \\
\addlinespace[2pt]
\bottomrule
\end{tabular}
\tablefoot{$D_{\rm KS}$ is the maximum absolute difference between the two
cumulative distributions, i.e., the peak of the residual
$\Delta$\,(FS $-$ non-FS) shown in the subpanels of Fig.~\ref{fig:QF_bins}.
Two-sided exact $p$-values, computed with \texttt{scipy.stats.ks\_2samp}.
None of the four comparisons rejects the null hypothesis of a common parent
distribution at the 95\% confidence level.}
\end{table}

\subsubsection{Color-mass distribution of galaxies}
\label{sec:color_magnitude}

We explored the galaxy populations of FSs and non-FSs in color--mass space. The top left and middle panels of Figure \ref{fig:color_mass} show  the galaxies' color-stellar mass distribution obtained when FSs and non-FSs are classified based on the $0.5R_{200}$ criteria. To compute the colors, we used the intrinsic galaxy $g$ and $r$ colors, extracted from the public IllustrisTNG dataset. As previously discussed in Sec. \ref{sec:Halo_selec}, to increase the number of structures analyzed in this section,  we further extended our sample by selecting structures at $z=0$, 0.2, and 0.4. Each distribution was normalized so that FSs and non-FSs had the same number of galaxies. The two distributions are globally similar, following the expected bimodal structure of a heterogeneous sample of groups and clusters, with a well-defined red sequence ($(g-r) \approx 0.7$) and a blue cloud. 

To further search for differences  between the internal distribution of galaxies of each subsample, in the right top panel we show the residual map obtained after subtracting the normalized number of FS and non-FS galaxies in each color-stellar mass bin. The residuals first reveal a clear overabundance of red galaxies in FSs, as well as a more populated blue cloud and green valley region in non-FSs. In addition, with respect to non-FSs, at $(g-r) \approx 0.6$ FSs show a deficit of high-mass galaxies ($10^{11}$ < M$_{\rm \star}$/M$_{\rm \odot}$ < 10$^{11.5}$). The deficiency of FS massive galaxies in the green valley  is  accompanied by an excess of high-mass galaxies (M$_{\rm \star}$/M$_{\rm \odot}$ > 10$^{11.5}$). This is better highlighted in the second row of panels, where we show stellar mass histograms of all galaxies with $(g-r) > 0.6$. The histogram are normalized by the total number of galaxies in each sample. Non-FSs show a $\approx 7 \%$   excess in this mass range. This difference is a direct consequence of the FS and non-FS structure selection criteria. For a structure to be classified as a non-FS, a massive luminous galaxy has to be found, in this case, within $R \le 0.5\ R_{200}$, ensuring an r-band magnitude gap (with respect to its BSG) $\Delta m_{1,2} < 2$ mag. In these models, these massive galaxies are preferentially located near the red sequence.  Very similar results were obtained when FSs and non-FSs were selected according to the $1\ R_{200}$ selection criterion (bottom panels). FSs show a more prominent red sequence and a less highly populated blue cloud than non-FSs. Although slightly less pronounced, we also found an overabundance of massive galaxies ($10^{11}$ < M$_{\rm \star}$/M$_{\rm \odot}$ < 10$^{11.5}$)  at $(g-r) > 0.6$. }

Our results agree well with the differences observed in $f_{\rm Q}$ between FSs and non-FSs discussed in Sect. {\ref{sec:fraction_of_quenched_galaxies}}, especially for the $0.5R_{\rm vir}$ selection criterion. Interestingly, A25 showed that structures in a dynamically disturbed state have a higher fraction of blue galaxies and a lower fraction of quenched galaxies than their relaxed counterparts, indicating a more recent  but short-lived star formation activity in disturbed structures than in those found within relaxed structures. We explore this further in the following section.

\begin{figure*}
\centering
\includegraphics[width=1\linewidth]{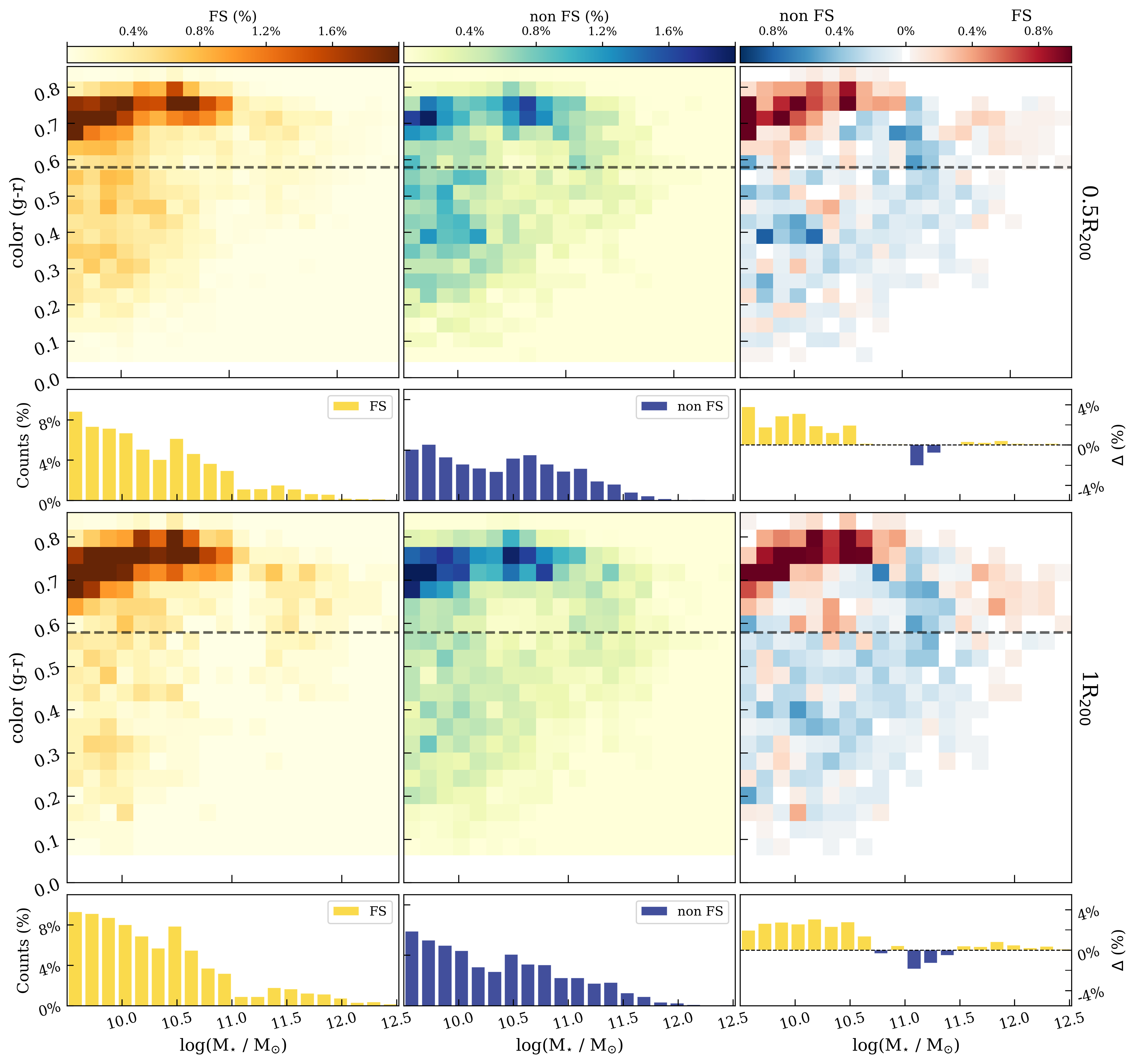}
    \caption{Hess diagrams showing the $(g-r)$ color vs. stellar mass distribution of satellite galaxies for FS (left column) and non-FS (middle column) populations and their residuals $\Delta$ (FS $-$ non-FS, right column). The systems are classified at $0.5\,R_{200}$ and $1\,R_{200}$ (upper and lower main panels, respectively), each accompanied by a marginal subpanel showing the stellar mass distribution in $(g-r) \geq 0.58$, marked by the dashed horizontal line. The analysis combines galaxy samples at $z = 0$, $z = 0.2$, and $z = 0.4$.}
\label{fig:color_mass}
\end{figure*}

\subsection{Dynamical state via gas-BSG centroid offsets}
\label{sec:dynamical_state}

Given the relaxed dynamical nature of fossil systems proposed in previous studies, and our result highlighting the absence of recent major mergers  (Sec.~\ref{sec:analysis:magnitudegap}) and a more dominant red sequence (Sec.~\ref{sec:color_magnitude}) in such structures, we compared the galaxy populations we obtained based on these two  classification schemes. 

First, instead of subdividing our structure sample based on the magnitude gap, we classified structures based on the shift between the BSG center of density and the gas center-of-mass (CoM). As described in Section~\ref{sec:methods:dynamical}, we separated the dynamical state of each system into relaxed and perturbed following the thresholds established by \cite{Zhang2022} and A25. Structures with $D_{\rm BSG-CM} > 0.4\ R_{\rm vir}$ and $D_{\rm BSG-CM} \le  0.1\ R_{\rm vir}$ were classified as dynamically perturbed and relaxed, respectively. We discarded structures with intermediate values of  $D_{\rm BSG-CM}$ from this analysis. We recall that to increase the number of analyzed structures, we selected systems at $z=0$, $z=0.2$, and $z=0.4$. This created a sample of $124$ relaxed and $78$ perturbed systems. 

\begin{figure*}
    \centering
    \includegraphics[width=1\linewidth]{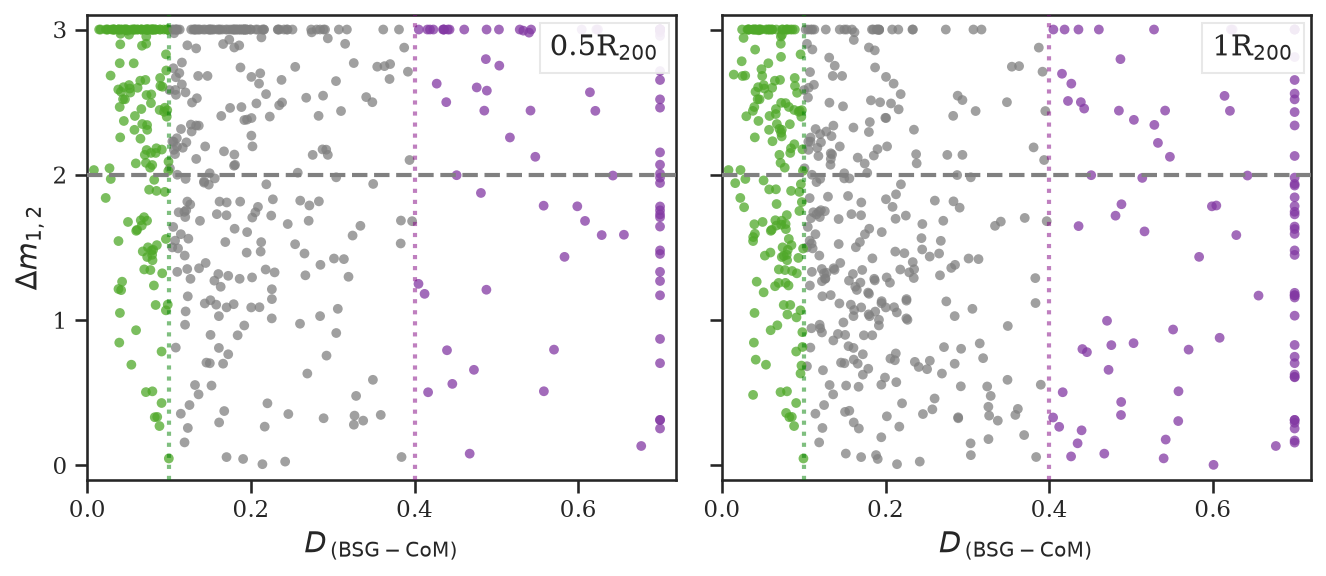}
    \caption{$\Delta m_{1,2}$ as a function of $D_\mathrm{BSG-CM}$ for the sample selected at 0.5R$_{200}$ (left) and 1R$_{200}$ (right). The green lines indicate the relaxed-structure threshold ($D_\mathrm{BSG-CM}$ < 0.1$R/R_{200}$), and the purple lines are associated with perturbed structures ($D_\mathrm{BSG-CM}$ > 0.4$R/R_{200}$). The dashed gray lines show the separation in magnitude between FSs and non-FSs. Values exceeding $\Delta m_{1,2} \geq 3$ and $D_\mathrm{BSG-CM} \geq 0.7$ are clipped at those limits.}
    \label{fig:scatter_dyn}
\end{figure*}

In Figure~ \ref{fig:scatter_dyn} we show the distribution of the magnitude gap $\Delta m_{1,2}$ as a function of the dynamical state parameter $D_{\rm BSG-CM}$ for our full sample of structures at $z=0$. The left and right panels correspond to the classifications performed within $0.5R_{200}$ and $1R_{200}$, respectively. In both panels, the horizontal dashed gray line marks the fossil threshold ($\Delta m_{1,2} = 2$), while the vertical dotted green and purple lines indicate the boundaries adopted to define relaxed ($D_{\rm BSG-CM} \leq 0.1$) and perturbed ($D_{\rm BSG-CM} \geq 0.4$) dynamical states, respectively. The two panels reveal a broad scatter in $\Delta m_{1,2}$ across the full range of $D_{\rm BSG-CM}$, suggesting that no simple one-to-one correspondence exists between the magnitude gap and the dynamical state of a structure. We quantified this via the Spearman correlation between $\Delta m_{1,2}$ and $D_{\mathrm{BSG-CM}}$. Within $0.5\,R_{200}$ (N $= 519$) we obtained $\rho = -0.10$ ($p = 0.02$) , and the coefficient within $1\,R_{200}$ (N $= 554$) is $\rho = -0.28$ ($p = 2\times 10^{-11}$). These two coefficients indicate a negligible and weak anticorrelation within 0.5 R$_{200}$ and 1 R$_{200}$, respectively. This confirm that the magnitude gap is a poor predictor of the dynamical state of an individual system. Nevertheless, a qualitative trend is apparent: systems with large magnitude gaps ($\Delta m_{1,2} \geq 2$) tend to preferentially cluster at low values of $D_{\rm BSG-CM}$, which is consistent with the expectation that fossil systems are more dynamically relaxed on average. Conversely, structures with small magnitude gaps are distributed across the full range of dynamical states, including relaxed and perturbed regimes. This behavior is consistent in both apertures, although the $R_{200}$ panel displays a somewhat more dispersed distribution, reflecting the inclusion of a larger satellite population at greater projected distances.

\begin{figure*}
    \centering
    \includegraphics[width=1\linewidth]{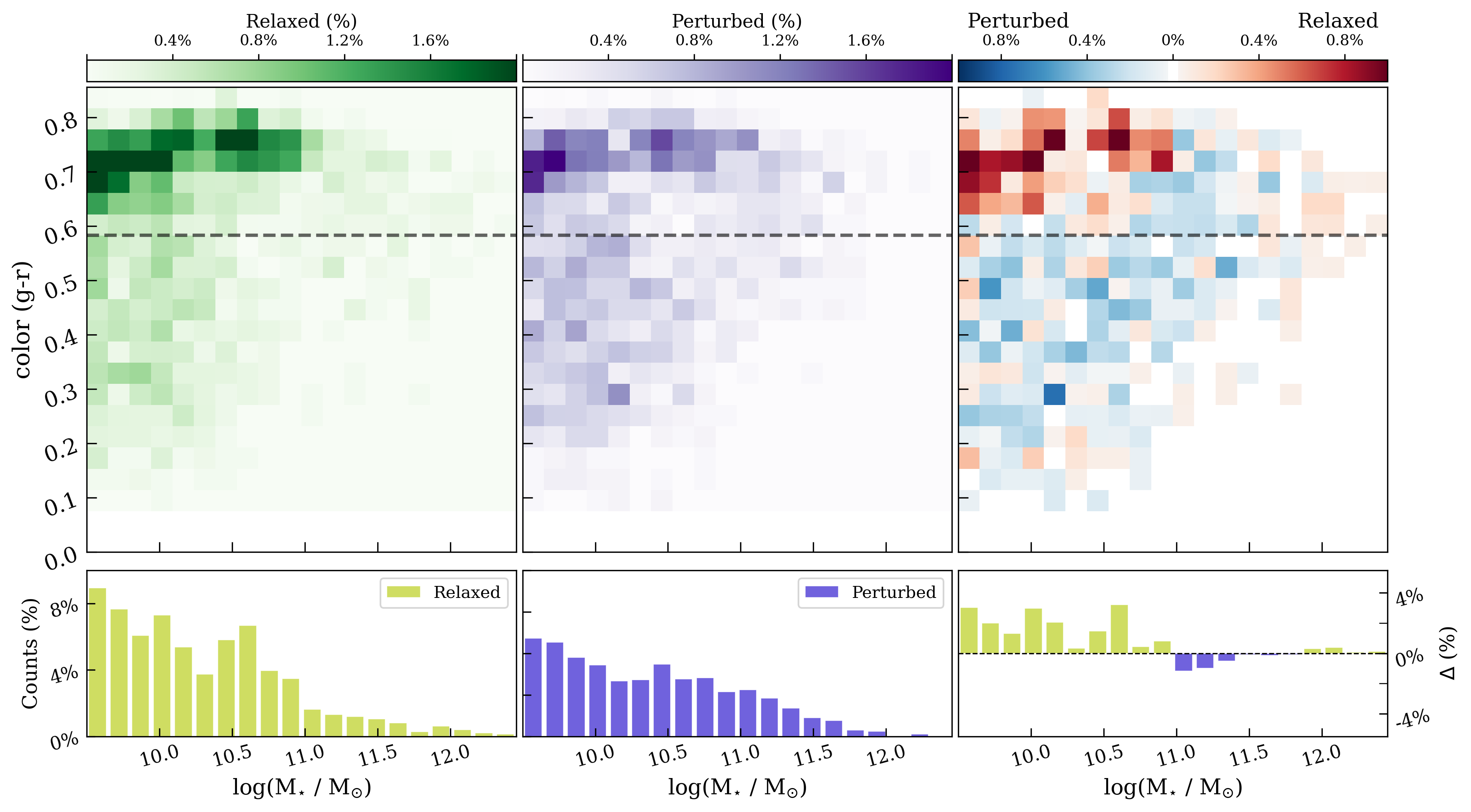}
    \caption{Hess diagrams showing the $(g-r)$ color vs. stellar mass distribution of satellite galaxies for relaxed (left column) and perturbed (middle column) systems, and their residuals $\Delta$ (relaxed $-$ perturbed, right column). The systems are classified as relaxed or perturbed based on the BSG--center-of-mass offset, with $D_{\rm BSG-CM} < 0.17\,R_{200}$ and $D_{\rm BSG-CM} \geq 0.17\,R_{200}$, respectively. Each main panel is accompanied by a marginal subpanel showing the normalized stellar mass distribution. The dashed horizontal line marks the red-sequence boundary at $(g-r) = 0.6$. The residual panel highlights an excess of red-sequence galaxies in relaxed systems and an overabundance in the blue cloud for perturbed systems. The analysis combines galaxy samples at $z = 0$, $z = 0.2$, and $z = 0.4$.}
    \label{fig:stack_dynamical_state}
\end{figure*}

Figure~\ref{fig:stack_dynamical_state} shows the color--stellar mass diagrams for relaxed (left) and perturbed (right) systems. These two types of systems are very similarly distributed, with a well-populated red sequence and a blue cloud. As before, to highlight the differences between these two system classes, we show the residual map in the right panel. Interestingly, and as previously discussed by A25, the red sequence is significantly more populated in relaxed clusters, while the remaining area, including the blue cloud, is more prominent in disturbed clusters. The comparison between the residual maps obtained from FSs and non-FSs (Fig. \ref{fig:color_mass}) and from disturbed and relaxed structures (Fig. \ref{fig:stack_dynamical_state}) reveals that  except for overabundance of galaxies at $10^{11}$ < M$_{\rm \star}$/M$_{\rm \odot}$ < 10$^{11.5}$ and $(g-r) \gtrsim 0.6$ in non-FSs,  the residual galaxy distribution is very similar. This result suggests a broad parallelism between the two classification criteria.  To further explore this, we computed $\Delta m_{1,2}$ for relaxed and disturbed structures. We found mean values of $\Delta^{\rm relaxed}_{m_{1,2}} = 2.32 \pm  0.76$ and $\Delta^{\rm disturbed}_{m_{1,2}} = 1.72 \pm 1.12$, indicating that the two classifications are reasonably consistent, but are not perfectly correlated. Even though the mean magnitude gap of the disturbed cluster population is below the chosen threshold, nearly 40\% of the disturbed structures have  $\Delta^{\rm disturbed}_{m_{1,2}} > 2$. On the other hand, $\approx 40$ \% of the relaxed cluster show a $\Delta^{\rm relaxed}_{m_{1,2}} < 2$.  We also calculated $D_{\rm BSG-CM}$ for all FSs and non-FSs. We focused on the fossil and non-fossil structure classification based on the canonical $0.5R_{200}$ criterion. Interestingly, the median values $D_{\rm BSG-CM}^{\rm FS} = 0.14 \ R_{200}$ and $D_{\rm BSG-CM}^{\rm non-FS} = 0.16 \ R_{200}$ are very similar, indicating that the two populations are in an intermediate dynamical state, rather than relaxed or disturbed. 

Our results suggests that the magnitude gap, while indicative of different assembly histories (e.g.,  Sec.~\ref{sec:analysis:magnitudegap}), does not necessarily imply a fundamentally different current dynamical state. FSs and non-FSs appear to be moderately perturbed systems that have not yet relaxed from their assembly histories. The slightly smaller offsets in FSs are consistent with their earlier cessation of major mergers (3--4 Gyr ago for FSs vs. ongoing for non-FSs), allowing marginally more time for partial relaxation, but this is insufficient to reach the relaxed regime as a population.

\section{Discussion}
\label{sec:discussion}

\subsection{Magnitude gap evolution and fossil fraction}

Early observational estimates that were derived in parallel with the first definitions of the magnitude gap gave fossil-structure fractions of 8--20\% relative to non-fossil systems within the same luminosity range \citep{Jones2003, Ebeling1997}. More recent surveys reported higher fractions: \citet{Gozaliasl2014} reported $22.2 \pm 6\%$
of fossil structures out of the total analyzed sample at $z \leq 0.6$ from
CFHTLS+XMM-LSS \citep[see also][]{Adami2020}. In addition, from SDSS-DR4 and RASS data, a number density of $2.83\times10^{-6}\,h^{3}\,\mathrm{Mpc}^{-3}$ was estimated \citep{LaBarbera2009, Vikhlinin1999}. A comparison between our results and those from previous observational studies reveals an excess of fossil systems in IllustrisTNG100. With our selection criteria set at $0.5R_{200}$, we found a fossil fraction of $\sim 68\%$ at $z=0$. When we extended the range to $1R_{200}$, the distribution became more balanced and the fossil fraction reduced to $\sim 48\%$. Very similar fractions were obtained when we restricted our structure sample to the low-mass range, that is, $10^{13} - 10^{13.5}$ $M_{\odot}$. The excess of fossil over non-fossil systems was also present when the sample was split into mass bins, showing an overabundance in simulations with respect to observations across the whole mass range \citep[e.g.][]{Dariush2007, Cui2011, Gozaliasl2014, Adami2020}. Observationally,
fossil systems remain a minority with respect to non-fossil systems at all masses, and
this minority becomes increasingly pronounced toward cluster scales; the disagreement with our results is therefore largest for the most massive structures (i.e., fossil clusters). This excess of fossil systems relative to observational surveys
\citep{review} is consistent with what was reported in other cosmological
simulations, such as K17. Using the Illustris simulation, K17 reported fossil fractions
of $\sim 80\%$ at $0.5R_{200}$ and $\sim 54\%$ at $R_{200}$ in Illustris for groups
with $M_{200}=10^{13}-10^{13.5}\,M_\odot/h$. The updated feedback model implemented in TNG-100 partially reduces the discrepancy with observations, but the fossil structure excess still remains. It is worth highlighting that we analyzed all structures at $z=0$. For structures at $z=0.4$ we found a fossil structure fraction of 52\% and 32\% for the $0.5R_{200}$ and $R_{200}$ criteria, respectively. This indicates that the fossil-structure fraction depends on $z$. This is relevant for a comparison with observational samples, which typically span a range of intermediate redshifts rather than $z = 0$. Furthermore, as in K17, our structures are selected by halo mass, whereas observational samples of fossil systems are typically selected by imposing a minimum X-ray luminosity \citep[e.g.][]{Jones2003}. Although the two selections are linked through the $L_{\rm X}$--$M$ scaling relation, the intrinsic scatter of this relation implies that a mass-selected sample is not strictly equivalent to an X-ray-selected one, so the comparison between the two sets of fractions is not straightforward. Finally, a bias in the classifications of FSs due to the lack of galaxy member confirmation through spectroscopic data can affect the observed fractions partially \citep[see e.g.][]{Santos2007, review}, which remains observationally challenging \citep{Oman2016}: \citet{Santos2007}, for instance, reported 34 fossil candidates lacking such confirmation. We will further explore this in a follow-up study. We note, in addition, that most simulation-based studies of fossil systems focused on their evolution \citep{Kanagusuku2016, Vonbendabeckmann2008}, formation \citep{Dariush2007, Cossairt2022}, or environment \citep{DiazGimenez2008, DiazGimenez2011}, without a control sample that enabled a direct comparison between the fossil and non-fossil fractions. This makes it necessary to characterize fossil systems as a population, relative to a non-fossil control sample, as we have done in this work.

Our analysis revealed that the magnitude gap in fossil structures emerged relatively recently in cosmic time, with FS and non-FS populations diverging approximately 3--6 Gyr ago, depending on the mass regime (Section~\ref{sec:analysis:magnitudegap}). Rather than confirming the traditional view that fossil systems represent the oldest, most evolved structures in the universe, our results suggest that the magnitude gap is primarily a tracer of recent assembly history. This interpretation is supported by our analysis of the BSG-to-satellite stellar mass ratio (Sect.~\ref{sec:analysis:magnitudegap}), which shows that FSs experienced significantly smaller recent accretion events than non-FSs. This suggests that the key distinction between FSs and non-FSs might lie not in when they formed \citep{review}, but in when they ceased major mergers. FSs appear to have exhausted their reservoir of massive satellites 3--4 Gyr ago, while non-FSs continue to accrete massive companions. This ongoing accretion in non-FSs prevents the establishment of a stable large magnitude gap by continuously introducing bright galaxies into the system.

Evidence of a fossil phase was reported by K17, supporting the results from \citet{Vonbendabeckmann2008}. They  reported that for the $R_{200}$ selection criteria, all $z=0$ FSs were previously non-FSs at higher redshifts, and vice versa. This was inferred from the behavior of the mean of $\Delta m_{1,2}$, $\overline{\Delta m_{1,2}}$ as a function of time. They found that FSs and non-FSs switched  their classification, as a population, at $z \approx 0.3$. In contrast, while we found that all FSs have gone through a non-FS phase (at $t_{\rm lb} \gtrsim 5$ Gyr), the non-FSs $\overline{\Delta m_{1,2}}$ remained below our threshold ($\Delta m_{1,2} < 2$ ) over the full considered period (9 Gyr). In other words, $z=0$ non-FSs, as a population, preserved their non-FS status since $z \approx 1$. 

The situation differs under the more traditional FS structure selection criterion based on $0.5R_{200}$. K17 found that for the FS and non-FS populations, $\overline{\Delta m_{1,2}} > 2$ over most of the last 10 Gyr of evolution. Present-day non-FSs only attained $\overline{\Delta m_{1,2}} < 2$ during the last 0.5 Gyr of evolution. Our results show a different scenario. For the low-mass regime, we found that the FS population experienced a non-FS phase at $t_{\rm lb} \gtrsim 5$ Gyr. In contrast, present-day non-FSs show $\overline{\Delta m_{1,2}} \lesssim 2$ over the full period. Nonetheless, this population shows large scatter in the time evolution of $\Delta m_{1,2}$, indicating that nearly all present-day non-FSs have undergone a transient fossil phase \citep[in agreement with][]{Vonbendabeckmann2008}. For higher-mass structures, we found for all structures $\overline{\Delta m_{1,2}} > 2$, indicating a common a fossil phase. However, in the models we considered, the non-FS population arises $\approx 2$ Gyr ago, earlier than in K17. As discussed above, this difference likely reflects the updated AGN feedback model, which prevents the BSGs runaway stellar mass growth seen in the original Illustris model.

\subsection{Global properties of the galaxy populations}

As previously discussed, when classifying structures using the traditional $0.5R_{\rm vir}$ criterion, we found that the present-day non-FS population arose $\approx 2$ Gyr ago. This is due to the accretion of massive galaxies or group structures that produce the required  $\Delta m_{1,2}$. The mild differences in the quenched fractions and color-mass distributions between FSs and non-FSs reported in Sect.~\ref{sec:fraction_of_quenched_galaxies} and Sect.~\ref{sec:color_magnitude}, namely a more prominent red sequence in FSs and an overdensity of massive green-valley galaxies in non-FSs associated with the secondary galaxies that drive the required $\Delta m_{1,2}$, agree well with \citet{Aldas2024}, who showed that during the early stages of significant accretion events and up to approximately 500Myr after infall, there is a period of enhanced star formation in the  associated galaxies. This SFR enhancement is followed by an SFR suppression (within 2 Gyr) as the cluster evolves toward a new equilibrium state. The short lifetime of these SFR enhancements explains the relatively mild differences in the $f_{\rm Q}$.   

Our results shows that no significant differences in the SF activity of FSs and non-FSs are expected. The quenching timescales in massive structures are typically $\lesssim 3 -4$ Gyr or shorter   \citep[][]{Wetzel2012,Walters2022, Pallero22}. Since non-FSs are associated with massive accretion events within $t_{\rm lb} \lesssim 2$ Gyr, most galaxies in a non-FS are expected to have already reached a fully quenched state. In addition, preprocessing effects, where satellites are quenched in smaller infalling groups before accretion onto the main halo, erase observable differences between FSs and non-FSs by the time satellites reach our selected radii \citep[e.g.][]{Pallero19, Pallero22}.  While a detailed analysis of preprocessing is beyond the scope of this work, these effects warrant further investigation in future studies that track satellites from first infall through their complete orbital histories.

\subsection{Dynamical state: Magnitude gap versus gas-BSG offset}

Our dynamical state analysis showed that while the magnitude gap separates systems by recent assembly history, it does not clearly separate them by current dynamical state\footnote{We recall that this classification is based purely on the shift between the BSG position and the gas CoM.}. FSs and non-FSs occupy an intermediate dynamical regime, with median gas-BSG offsets of $D_{\rm BSG-CM} \approx 0.15$, which is well above the relaxation threshold of 0.1 $R_{200}$, but also well below the disturbed threshold of 0.4 $R_{200}$ \citep{Zhang2022, Aldas2024}.  Even when selecting FSs with $\Delta_{m_{1,2}} > 3$, we found a median $D_{\rm BSG-CM} \approx 0.15$.
We reach a similar conclusion when we directly classified structures by $D_{\rm BSG-CM}$: the mean magnitude gaps of the relaxed and disturbed populations (Sect.~\ref{sec:dynamical_state}) are reasonably consistent but not perfectly correlated with the classification into FSs and non-FSs.
This finding has important implications for interpreting the magnitude gap as a dynamical proxy. Previous studies have often assumed that systems with large magnitude gaps are dynamically relaxed, having had sufficient time for the luminosity hierarchy to develop through completed mergers and settled potential wells \citep{Dariush2007, Gozaliasl2014}. Our results suggest that this assumption requires refinement: while FSs do show modestly smaller gas-BSG offsets than non-FSs, indicating that their earlier cessation of major mergers has allowed for the partial relaxation of the core regions, neither population appears to have reached a relaxed state. The FS and non-FS populations maintain average offset values characteristic of moderately perturbed systems, suggesting that complete dynamical settling remains an ongoing process even in systems with stable magnitude gaps. Several studies currently incorporate several different indicators to isolate relaxed from perturbed structures \citep[e.g.][]{Hyowom2025, VelizAstudillo2025} 

\citet{FOGO7} reported that substructures were detected around fossil systems when the analysis extended beyond $0.5R_{200}$ and that their presence modifies the magnitude gap when measured out to $2R_{200}$. These substructures perturb the outskirts of the host system without directly affecting the magnitude gap, consistent with the lack of a strong correlation between $\Delta m_{1,2}$ and the BSG-CM offset reported above.

When examining the galaxy color-stellar distributions of structures classified by their dynamical state, rather than fossil state (Figure~\ref{fig:stack_dynamical_state}), we recovered the results presented in \citet{Aldas2024}. Systems classified as relaxed show a marked preference for red galaxies compared to more perturbed systems, consistent with the fact that galaxies in structures with recent massive accretion events experience a short period of enhanced SFR and, thus, bluer colors. The differences are similar to those found between FSs and non-FSs. However, the green-valley overdensity observed in non-FSs is absent in the disturbed subsample (with respect to relaxed systems). This highlights stricter conditions behind the non-FS classification, that is, the need for an extremely bright galaxy within 0.5 $R_{200}$.

\section{Summary and conclusions}
\label{sec:conclusions}

We have used the IllustrisTNG simulations to characterize the emergence of FSs and non-FSs. With respect to the original Illustris simulation, the updated AGN feedback model implemented in IllustrisTNG results in a better mass / luminosity determination for the central BSGs, allowing for smaller magnitude gaps to arise. This has a significant effect when characterizing fossil structures. One of our goals was to understand whether this classification can be used to discern between relaxed and disturbed structures. To this end, we followed the time evolution of the selected structures magnitude gap and quantified and compared the global properties of their galaxy populations. We also classified structures as dynamically relaxed and perturbed based on the shift between their CoM and the location of their BSG. Our main results are summarized below.

\begin{itemize}
    \item The fraction of fossil systems we found is smaller by $\approx 20 - 30\%$ than that presented in K17 using the Illustris models. Nonetheless, we still found an overabundance of FSs with respect to observations. Other aspects contributing to this overabundance, such as the lack of X-ray luminosity criterion to select structures, must be explored.
    
    \item The evolution of $\Delta \text{m}_{1,2}$ over time depends on the halo mass. High-mass halos develop magnitude gaps earlier and exhibit lower dispersion than low-mass halos.
    
   \item All present-day FSs were non-FSs at higher redshift, supporting the fossil phase scenario proposed in previous studies.

    \item Compared to non-FSs, FSs experienced fewer massive last satellite accretion events in the last 6\,Gyr  ($\mu_{\star}^{\text{FS}} = 0.17$ vs.\ $\mu_{\star}^{\text{NFS}} = 0.39$). The magnitude gap is mainly a tracer of the recent accretion history. 
    
    \item The distribution of the quenched fraction ($f_{\text{Q}}$) differs in the halo mass bins. Low-mass halos show a wide variety in $f_{\text{Q}}$, including $\sim 20\%$ of the structures that are dominated by star-forming galaxies, while high-mass halos are typically characterized by higher $f_{\text{Q}}$ values.

    \item When the fossil condition is determined based on galaxies located within $0.5R_{200}$, the residuals of the $f_{\text{Q}}$ distribution also reveal differences between FSs and non-FSs. A KS test yielded inconclusive results, likely due to the limited sample size. Nonetheless, this results suggest slightly larger fractions of star-forming galaxies in non-fossil structures. 

    \item A comparison of the color–mass distribution of FSs and non-FSs revealed a depletion of high-mass satellites in the green valley for FSs. This depletion is a direct consequence of the need for a secondary massive and bright galaxy in non-FSs, which serves to guarantee a small magnitude gap. Consistent with this picture, we found the red sequence to be more prominent in FSs, while the blue cloud is more populated in non-FSs.
    
    \item As in previous studies, we found that in color--mass space, relaxed clusters dominate the red-sequence region, while perturbed clusters show an overabundance in the blue cloud. In addition to the overdensity (underdensity) of massive galaxies in the green valley of non-FSs (FSs), FSs (non-FSs) and relaxed (disturbed) structures show a similar color-stellar mass distribution.

\end{itemize}

The comparison between fossil classification (based on the magnitude gap) and dynamical classification (based on the BSG--gas centroid offset) revealed that these two diagnostics probe different aspects of system evolution. The magnitude gap effectively separates systems by their recent merger histories, as reflected in the significantly lower $\mu_{\star}$ values of FSs over the past 6\,Gyr. However, it does not clearly separate systems by their current dynamical state: FSs and non-FSs span a range of relaxation levels, and neither population is relaxed on average. This is likely a consequence of the sensitivity of the magnitude gap to the selection of satellites within different radii, which further limits its ability to capture a system's global dynamical properties. Furthermore,  ongoing low-mass satellite accretion is likely to also play a role,  modifying the gas distribution even in systems with large magnitude gaps. 

Future work should focus on extending this analysis to better characterize the dynamical evolution of fossil systems and the processes that shape their current state. In particular, the stellar content of the intracluster light and the BSG offers a promising avenue: studying their evolution jointly and separately might disentangle the relative contributions of major mergers, tidal stripping, and minor accretion to the buildup of the central mass reservoir. The intracluster-light fraction and its radial profile are expected to retain imprints of the assembly history, and comparing these between FSs and non-FSs might provide independent constraints on the merger timescales suggested by the magnitude gap. Complementary dynamical 
diagnostics beyond the BSG--gas centroid offset, such as the shape of the velocity dispersion profile or the degree of substructure in the satellite phase space, might further clarify the relation between fossil classification and the true relaxation state of these systems.

\begin{acknowledgements}

MVS thanks their co-authors for helpful discussion, comments, and patience throughout this work. MVS also thanks A. Monachesi for insightful comments and suggestions. This research was supported by the National Agency for Research and Development (ANID) through the Scholarship Program DOCTORADO BECAS CHILE/2021 – 21211323. FAG and MVS acknowledge support from the ANID BASAL project FB210003, the ANID FONDECYT Regular grant 1251493 and from the HORIZON-MSCA-2021-SE-01 Research and Innovation Programme under the Marie SklodowskaCurie grant agreement number 101086388. F.A. acknowledges financial support from ANID through the FONDECYT Postdoctoral Fellowship, Grant No. 3260589. FA  acknowledges support by the Italian Research Center on High Performance Computing, Big Data and Quantum Computing (ICSC), project funded by European Union - NextGenerationEU - and National Recovery and Resilience Plan (NRRP) - Mission 4 Component 2, within the activities of Spoke 3, Astrophysics and Cosmos Observations; by the PRIN 2022 PNRR project (202259YAF) “Space-based cosmology with Euclid: the role of High-Performance Computing.”; by the PRIN 2022 (20225E4SY5) “From ProtoClusters to Clusters in one Gyr”; by the INAF Astrofisica Fondamentale Large Grant 2023 “Witnessing the Birth of the Most Massive Structures of the Universe”, and partial financial support from the INFN Indark Grant. D.P. acknowledges support from the N\'ucleo Milenio de Galaxias (MINGAL), ANID - MILENIO NCN202\_112. We are grateful to the IllustrisTNG collaboration for making their simulations publicly available.
We thank the anonymous referee for useful comments and suggestions that helped improve the manuscript. The authors also acknowledge the use of Claude AI (Anthropic) for language editing and improving the clarity of the text.

\end{acknowledgements}

\bibliographystyle{bibtex/aa}
\bibliography{bibtex/refers}

\end{document}